\documentclass[]{spie}  %>>> use for US letter paper
\usepackage{amsmath,amsfonts,amssymb}
\usepackage{graphicx}
\usepackage[colorlinks=true, allcolors=blue]{hyperref}

\title{Deep and shallow data science for multi-scale optical neuroscience}
\author[a]{Gal Mishne}
\author[b]{Adam Charles}
\affil[a]{Halıcıoğlu Data Science Institute, Department of Electrical and Computer Engineering and the Neurosciences Graduate Program, UC San Diego, 9500 Gilman Drive, La Jolla, CA 92093 USA}
\affil[b]{Department of Biomedical Engineering, Kavli Neuroscience Discovery Institute, Center for Imaging Science, Department of Neuroscience, and Mathematical Institute for Data Science, Johns Hopkins University, Baltimore, MD 21287 USA}

\authorinfo{Further author information: (Send correspondence to G.M. and A.S.C.)\\G.M.: E-mail: gmishne@ucsd.edu\\A.S.C.: E-mail: adamsc@jhu.edu}

\begin{document} 
\maketitle

\begin{abstract}
Optical imaging of the brain has expanded dramatically in the past two decades. New optics, indicators, and experimental paradigms are now enabling in-vivo imaging from the synaptic to the cortex-wide scales. To match the resulting flood of data across scales, computational methods are continuously being developed to meet the need of extracting biologically relevant information. 
In this pursuit, challenges arise in some domains (e.g., SNR and resolution limits in micron-scale data) that require specialized algorithms. These algorithms can, for example, make use of state-of-the-art machine learning to maximally learn the details of a given scale to optimize the processing pipeline. In contrast, other methods, however, such as graph signal processing, seek to abstract away from some of the details that are scale-specific to provide solutions to specific sub-problems common across scales of neuroimaging. Here we discuss limitations and tradeoffs in algorithmic design with the goal of identifying how data quality and variability can hamper algorithm use and dissemination.  
\end{abstract}

\keywords{fluorescence microscopy, calcium imaging, functional imaging, data analysis}

\section{INTRODUCTION}
\label{sec:intro}  % \label{} allows reference to this section

Data science, in the form of statistical modeling and image processing, has become a central topic in extracting the most out of complex imaging data, such as taken by the latest in optical designs~\cite{benisty2022review}. Computational approaches in particular have blossomed as imaging methods have extended across the micro- and macro- scales that aim to glean information about how the brain processes information at all scales~\cite{charles2020toward}. These methods fundamentally take into account---implicitly or explicitly---properties of the data imbued by the imaging target and the imaging device. Colloquially we refer to these as the likelihood and the prior, after the successful terminology from the Bayesian inference and estimation literature. 

Due to the enormous range of targets (different brain areas across species) at different scales (cell bodies, widefield, dendrites, axons, vasculature) that exhibit different temporal dynamics (different indicators of calcium and voltage, hemodynamics, long-term AMPA receptor changes, etc.), fundamentally different tools have been developed by the community.
%across the literature.  
%GSP vs NNs
Some of these tools have aimed to solve core processing challenges, for example, denoising or identifying functional components in videos, across many different imaging contexts. Other tools instead have focused on concentrating effort toward solving a very specific set of challenges in a single context. As an extreme example of the latter, computational imaging techniques use specialized algorithms to process and interpret data from one specific instance of an imaging device. 

Inherent to these methods are thus a set of tradeoffs surrounding the ability of these algorithms to generalize across contexts. These tradeoffs can impact how new methods can be robustly assessed and shared. We discuss here a number of considerations one should consider both in the development of such algorithms and in the deployment and dissemination thereof.  

%and sometimes same tool can bridge across scales and morpho and data source

%data science for processing (denoising), inference and data generation and model validation 

\section{The promise of deep learning?}
One increasingly popular group of approaches in the analysis of functional optical imaging is to leverage advances in deep learning, where deep learning has been integrated into various points of the processing pipeline for functional imaging, from preprocessing, through inference to postprocessing, which we briefly review below.

% preprocessing - denoising 
One challenge deep learning methods have sought to address is denoising of the acquired video.
Functional imaging suffers from imaging-dependent noise, where signal-to-noise (SNR) can vary across different neuronal components due to differences in the expression of the indicator.
Identifying neural components, e.g., cell bodies, is enabled because the location of cells is fixed in the field-of-view and their activity results in fluorescence persisting over multiple frames.  
Extraction of spatial components and their corresponding time traces can be improved by incorporating a denoising step as part of preprocessing (following motion correction).
Recently, general advances from deep learning~\cite{lecoq2021removing,li2021reinforcing,weigert2018content} have been proposed for denoising calcium imaging.
One class of methods, including DeepInterpolation~\cite{lecoq2021removing}, SUPPORT~\cite{eom2023statistically} and DeepVID~\cite{liu2023two,platisa2023high}, leverage the spatio-temporal consistency of the fluorescence signal to separate it from the noise. In essence, thse methods train a neural network to predict a movie frame (for DeepInterpolation) or pixel (for SUPPORT) based on previous and past frames or the surrounding pixels in space and time, respectively. As independent noise cannot be predicted from the spatio-temporal neighborhood, it is thus effectively removed. 
A broader class of image-restoration methods are based on more traditional static architectures. CARE~\cite{wei2020comparison}, and the more recetnt XTC that specializes for sypatic imaging~\cite{xu2023cross}, are based on a U-net architecture trained on high- and low-resolution images. These images can be obtained by more expensive measurements in vivo, or as in XTC, by creating training data ex-vivo using super-resolution microscopy.

%Denoising approaches make different noise and signal model assumptions and should be used judiciously. 

The fundamental step for the analysis of functional imaging data is identifying spatial regions of interest (ROIs) and their corresponding time traces. 
While earlier methods mostly relied on regularized matrix decomposition~\cite{pachitariu2016suite2p,giovannucci2018caiman,mishne2019learning,charles2021graft,mukamel2009automated}, recently, general deep learning methods for image segmentation have been adapted to the ROI extraction problem~\cite{klibisz2017fast,ronneberger2015u,apthorpe2016automatic,soltanian2019fast,bao2021segmentation,kirschbaum2020disco,stringer2021cellpose}. 
These are supervised approaches that require labeled training data in order to train a network to identify ROIs in new datasets. 

Some networks tackle only the problem of identifying spatial components and ignore the temporal dimension. 
UNet2DS~\cite{klibisz2017fast} is based on the fully convolutional UNet~\cite{ronneberger2015u} model, which was developed for biomedical image segmentation.  
UNet2DS takes the mean image as an input and outputs the probability of a pixel belonging to a cell or background. 
UNet2DS, however, ignores important temporal information, leading to a difficulty in separating overlapping neurons. Furthermore, it is dependent on the summary image that may be biased towards cells that fire more often, or have high expression despite exhibiting no activity. Similar to UNet2DS, Cellpose~\cite{stringer2021cellpose} is similarly based on a UNet architecture, with an added global ``style'' designator to help differentiate different image classes, and so suffers the same drawbacks. %Such methods also cannot differentiate active cells, i.e., cells that exhibit deviations from baseline fluorescence consistent with spiking events, from non-active cells.

Other networks have added temporal statistics or activity to their design~\cite{apthorpe2016automatic,soltanian2019fast,bao2021segmentation}. 
A (2+1)D convolutional neural network~\cite{apthorpe2016automatic} was trained on spatiotemporal sliding windows and the output represented the probability of a pixel belonging to an ROI centroid. 
This approach better suppressed noisy detections compared to a 2D network that only took as input the time-averaged image, yet still only provided spatial components as outputs.

STNeuroNett~\cite{soltanian2019fast} is a 3D convolutional neural network for neuron segmentation, trained on overlapping spatiotemporal blocks and outputs a binarized probability map. Shallow U-Net Neuron Segmentation (SUNS)~\cite{bao2021segmentation} simplifies STNeuroNet's architecture for speed and incorporates temporal information by applying a temporal matched filter and whitening to the input video. 
Both methods post-process the output probability maps by clustering to detect individual cells and aggregating detected cells across the maps.
The design of these networks incorporates temporal information while efficiently reducing it in order to obtain fast processing speeds and low-latency inference in closed-loop experiments. 
Finally, DISCo (Deep learning, Instance Segmentation, and Correlations)~\cite{kirschbaum2020disco} uses a combination of time-trace correlation-based pixel segmentation on a graph and a convolutional neural network to identify individual spatial profiles. 
While these these methods incorporate temporal information in different ways, they only identify cells and do not address the issue of estimating time traces.

% postprocessing
Finally, deep learning has also served in postprocessing to classify identified regions-of-interest (ROIs) as either true or false, based on human-annotated training data, to improve the precision of calcium imaging analysis methods, e.g., in CaImAn~\cite{giovannucci2018caiman}.

%%%%%
On the practical side, training these networks might involve a lot of training data and time.
The gain is the fast processing speed of these networks at test time.
Critical to their success is 1) the availability of training data, 2) sufficient computing resources, and 3) the generalization of the trained system to new datasets.
These impact the performance of deep learning models when deployed to unseen data in a new lab not encountered in training. 
As with general deep learning models, the exact expected biases of the black box system are unknown.

 % on their own data that would probably work well for data they are collecting - good local solution)
%- many labs have their own annotated data (either from other methods or manual labeling and can train networks on their own data that would probably work well for data they are collecting - good local solution)
%- use of neurofinder: temporal statistics might be outdated (older indicators, slow framerates), mislabeled data

\section{10 labs, 11 opinions: the out-of-distribution problem.}

One of the most frustrating situations is the inability of an algorithm that worked fantastically on one dataset but failed utterly on another. In neuroimaging---wherein surgical protocols, custom microscopes, and variability in environment can impact signal statistics from lab-to-lab---this effect is further frustrated when a method thoroughly validated on many datasets from one lab do not perform to the same degree in another lab. One reason can be traced back to an insidious type of overfitting to the distribution of one lab's data, while even a similar lab might have data that may as well be \emph{out of distribution (OOD)}.

OOD is a central challenge in machine learning more generally~\cite{geisa2021towards,fort2021exploring}. In particular, the fact that much of modern-day machine learning focuses on deriving much of what can be derived from the data itself, a central axiom is that the full dataset follows a static distribution. When, of the full distribution (say all videos of calcium imaging of GCaMP-6f transgenic mice in Hipocampus area CA1), only a biased subset is used to train the ML algorithm (say all videos of calcium imaging of GCaMP-6f transgenic mice in Hipocampus area CA1 taken at Johns Hopkins University) then the extra assumption is needed that the subset of the data follows the same distribution of the full data. If it does not then a different subset (say all videos of calcium imaging of GCaMP-6f transgenic mice in Hippocampus area CA1 taken at UC San Diego) will fail to leverage what the ML algorithm learned from the first subset. This effect is being continuously studied in more complex, modern, ML systems, however will continue to have an impact in how we can expect ML solutions to generalize across users. 

%Generalization is tougher to ensure, as image statistics can affect deep learning systems in a myriad of unexpected ways. Therefore extensive experimentation is required, for example by testing a trained system on imaging from different depths to explore the effect of tissue distortion~\cite{soltanian2019fast}.

One direct consequence of the ``every lab is OOD'' problem is that algorithms that learn too much from data can suffer drawbacks in dissemination. An obvious case is the use of deep learning. Deep learning fits highly over-parameterized models and is known to be able to interpolate within a distribution extremely well, while extrapolate very poorly to new domains. Thus, despite the fact that pre-trained networks are fast to run, OOD create a high level of uncertainty in using pre-trained networks accurately. 
In developing methods for functional imaging analysis, and especially for training deep learning models during method development, most use NeuroFinder~\cite{berens2017standardizing} and/or the Allen Institute Mouse Brain Observatory~\cite{Allen2019url}. However, both datasets suffer from both mislabeled data and missing labels, or erroneous time-trace estimates~\cite{gauthier2018detecting}.
Thus limited and mislabeled training data creates a limited and potentially distorted \emph{in distribution} that the deep learning model learns, which can affect performance in unexpected ways.
Not all is lost, however, as fine-tuning models on new datasets can be a work-around for new data sets, albeit at the cost of further annotation or data collection. 
In fact, many labs have manually labeled data acquired from their own optical systems and in the brain regions they typically study, this barrier can be somewhat mitigated provided easy processes by which to use this data to fine-tune the model.

Thus, even though pre-trained networks are fast to run, these challenges create a high level of uncertainty in using pre-trained networks accurately. 
Individual labs can instead choose to annotate their own data to train networks from scratch, such that they are optimized for the imaging used locally. While this solves some of the aforementioned challenges (assuming care is taken in annotation), the training procedure can be computationally intensive relative to a lab's computational capabilities.

Some methods explicitly seek to transcend such challenges by providing models that learn from data in an instance-specific setting. For example, matrix factorization and other such methods for source identification use more general assumptions of low-rank and non-negativity rather than learning full feature spaces as in deep learning. Danger, however, lies here as well. These methods also have multiple (numerous) parameters that are often fit in certain domains, or have regularizations that are fixed to work well for certain imaging contexts. For example, algorithms that constrain components to be contained in a small area in the image work excellently for somatic imaging across brain areas and species but fail for different imaging scales and non-compact or non-continuous morphologies (e.g., dendrites or widefield). 

\section{One pipeline does not fit all.}

The generalization constraints discussed create a fundamental foil in the field's goal of creating highly disseminable and reproducible data pipelines that can work ``out of the box''. Such a goal is highly appealing, as we should seek as much as possible to leverage the advances of others to optimize our data analysis. To realize this goal, however, two challenges must be met: robust assessment, and adaptive retraining.

The first of these challenges must face the fact that OOD means that the assessment performed of the pipeline and its constituent parts might not be correct for all labs/users. To enable new users to correctly choose an optimal computational approach, new approaches to validation must be developed. These validation approaches must promote data diversity and scale, rather than dissecting singular datasets. Promising avenues in this direction include the growing repositories of data in, e.g., the Dandi archive, and new simulation-based data generators~\cite{song2021neural}. 

The second challenge lies in having the pipeline meet the users. Currently, the goal of robust dissemination has been to make the code as immutable as possible~\cite{abe2022neuroscience}. In some dimensions, this is the correct requirement, as one does not want the code itself to change. In others, the pipeline must be adaptive to identify when the approach is failing at processing a dataset and which modification must be taken to improve the approach. Simple versions of such methods would be simple parameter re-fitting (e.g., using BayesOpt) or fine-tuning of deep learning models~\cite{pachitariu2022cellpose}. More complex solutions should be able to swap and reorder stages in the pipeline. None of these is currently ``out of the box'' for users and require significant manual intervention to adapt pipelines. 

\section{Conclusions}

The pace of neuroimaging analysis is rapid and unrelenting. Progress is often achieved within the context of specific labs tackling problems which, while critical to their work, would be invaluable to the community at large. Robust dissemination is thus key and must account for the OOD problem. 

New efforts offer promise. Efforts at the Allen Institute, the International Brain Lab, DataJoint, and NeuroCAAS, amongst others, are striving to create stable pipelines may lead to easier access. These efforts can form central points for tools that assess and correct OOD problems as new users build up their neuroimaging needs.

\acknowledgments % equivalent to \section*{ACKNOWLEDGMENTS}       
 G.M. was partially funded by a  Kavli Institute for Brain and Mind (KIBM) Innovative Research Grant award. We would like to thank Michael Xie for helpful feedback.

% References
\bibliography{report} % bibliography data in report.bib
\bibliographystyle{spiebib} % makes bibtex use spiebib.bst

\end{document}